\documentclass[prd,aps,showpacs,tightenlines]{revtex4}  %%preprint,
\usepackage{mathrsfs}
\usepackage{amsmath}
\usepackage{amssymb}
\usepackage{epsfig}
\usepackage{graphicx}
\usepackage{booktabs}
\usepackage{multirow}
\usepackage{subfigure}
\usepackage{color}
\begin{document}
%%%%%%%%%%%%%%%%%%%%%%%%%%%%%%%%%%%%%%%%%%%%%
%\renewcommand{\arraystretch}{0.5}
\newcommand{\psl}{ p \hspace{-1.8truemm}/ }
\newcommand{\nsl}{ n \hspace{-2.2truemm}/ }
\newcommand{\vsl}{ v \hspace{-2.2truemm}/ }
\newcommand{\epsl}{\epsilon \hspace{-1.8truemm}/\,  }

%%%%%%%%%%%%%%%%%%%%%%%%%%%%%%%%%%%%%%%%%%%%%%%%

\title{$P$-wave contributions to $B\to\psi\pi\pi$ decays in perturbative QCD approach }
\author{Zhou Rui}\email[]{jindui1127@126.com}
\affiliation{College of Sciences, North China University of Science and Technology, Tangshan, Hebei 063210,  People's Republic of China}
\author{Ya Li}\email[]{liyakelly@163.com}
\affiliation{Department of Physics, College of Science, Nanjing Agricultural University, Nanjing, Jiangsu 210095, People's Republic of China}
\author{Hsiang-nan Li}\email[Corresponding  author: ]{ hnli@phys.sinica.edu.tw}
\affiliation{Institute of Physics, Academia Sinica, Taipei, Taiwan 115, Republic of China}

\date{\today}
\begin{abstract}
We present the differential branching fractions for the
$B\rightarrow \psi \pi\pi$ decays with the charmonia $\psi=J/\psi,\psi(2S)$
in the invariant mass of the $P$-wave pion pairs in the perturbative QCD
approach. The two-pion distribution amplitudes (DAs) corresponding to both
longitudinal and transverse polarizations are constructed to capture important
final state interactions in the processes.
The timelike form factors, normalizing the two-pion DAs,
contain contributions from the $\rho$ resonance and radial excitations
fitted to the $BABAR$ $e^+e^-$ annihilation data.
Given the hadronic parameters for the two-pion DAs associated with the
longitudinal polarization which were determined in our previous study, and
tuning those associated with the transverse polarization, we accommodate
well the observed branching ratios and polarization fractions of the
$B\rightarrow J/\psi \pi\pi$ decays. Our predictions for the
$B\rightarrow \psi(2S) \pi\pi$ modes from the same set of parameters can
be tested in future LHCb and BelleII experiments. We also investigate
the sources of theoretical uncertainties in our calculation.
\end{abstract}

%\keywords{ }

\maketitle
%%%%%%%%%%

\section{Introduction}

The $B_{(s)}$ meson decay chains with charmonia and pion pairs in final states,
providing rich opportunities to search for intermediate resonances, have caught
both experimental and theoretical attention. The neutral and charged
$B\rightarrow J/\psi \pi\pi$ modes, first observed by the BaBar
Collaboration \cite{prl90091801,prd76031101}, may involve the
$J/\psi \pi$ and $\pi\pi$ intermediate channels.
No obvious exotic structures were found through the former,
and a series of resonant and nonresonant components with different
$\pi\pi$ invariant masses has been extracted though the latter
in the LHCb experiment \cite{prd87052001}.
Recent LHCb data \cite{prd87052001,prd90012003} have indicated
that the $B\rightarrow J/\psi \pi^+\pi^-$ decay spectrum is well
described by six resonances in the $\pi^+\pi^-$ channel,
$f_0(500)$, $\rho(770)$, $\rho(1450)$, $\rho(1700)$, $\omega(782)$, $f_2(1270)$,
with $\rho(770)$ being the dominant component, and that there is no evidence
for $f_0(980)$ production. The corresponding $B_s\rightarrow J/\psi \pi^+\pi^-$ decay
%corresponding to the exchange of the spectator quarks $d\leftrightarrow s$
can be described by an
interfering sum of five resonances, $f_0(980)$, $f_0(1500)$, $f_0(1790)$, $f_2(1270)$,
and $f'_2(1525)$ \cite{prd86052006,prd89092006}, among which the $S$-wave
$f_0(980)$ is the largest component \cite{prd86052006,prd89092006,prd79074024},
and the $D$-wave ones amount only up to a few percents. Because the $s\bar{s}$
pair produced in this mode is an isoscalar ($I=0$), it must form a zero isospin
meson, and $P$-wave resonances, such as the isovector $\rho(770)$, are forbidden.
The resonance structures in the $B_{(s)}$ meson decays into $\psi(2S)$ have not
been analyzed in detail due to a limited number of events \cite{npb871403}.

On the theoretical side, order-of-magnitude estimations for the rates of the above
modes have been performed in the chiral unitary approach \cite{prd90114004},
where a $B_{(s)}$ meson decay amplitude is followed by hadronization of
a quark-antiquark pair into two mesons and their further rescattering.
Given the input from a well-measured intermediate channel, the others
can be derived via their relations to the input one under the above
rescattering picture, and were found to compare reasonably well with
present data. The authors in Ref. \cite{ps88035101} calculated the
$B\rightarrow J/\psi \pi\pi$ branching ratios in the generalized factorization
and improved QCD factorization approaches, where the $\rho$ intermediate
resonance was described by a Breit-Wigner (BW) propagator. More recently,
final state interactions in the $B_{(s)}\rightarrow J/\psi \pi\pi$ decays
were extracted from data in a framework based on dispersion theory \cite{jhep02009}.
These works mainly focus on the $\rho(770)$ contribution to the $P$-wave
di-pion system, with the two radial excitations $\rho(1450)$ and $\rho(1700)$
and the $\rho$-$\omega$ interference being neglected. As stressed in
Ref. \cite{epjc3941}, the contributions from the two excited $\rho$ states to the
timelike pion form factor are indispensable, if one intends to accommodate the
measured space-like pion form factor from the timelike one through analytic continuation.
Several collaborations \cite{prd86032013,plb527161} have also successfully
fitted the $e^+e^-\rightarrow \pi^+\pi^-$ cross section in the vicinity
of the $\rho(770)$ resonance, with a small but clearly visible $\omega$-meson admixture.

It has been argued \cite{plb561258} that the dominant kinematic region for three-body
$B$ meson decays is restricted to the edges of a Dalitz plot, where two of the
three final state mesons form a collimated pair in the rest frame of the $B$ meson.
In this region, the proof of the corresponding factorization theorem is basically
similar to that for the two-body cases \cite{beneke,npb899247,jhep10117}. Hence,
the perturbative QCD (PQCD) approach \cite{prl744388,plb348597} is applicable to
three-body $B$ meson decays, albeit the underlying $k_T$ factorization has not been
proven rigorously \cite{prd88114014,prd94114014}. With the introduction of two-hadron
distribution amplitudes (DAs) \cite{G,DM,Diehl:1998dk,MP} to absorb the final state
interaction involved in the meson pair, the factorization formalism can be greatly
simplified. The factorization theorem holds for $B$ meson decays containing charmonia
in the heavy quark limit under the power counting specified in \cite{Kurimoto:2002sb}.
As a result, a typical amplitude for the $B\rightarrow \psi \pi\pi$ decays,
$\psi=J/\psi,\psi(2S)$, is written as \cite{plb561258}
\begin{eqnarray}
\mathcal{A}=\Phi_B \otimes H\otimes \Phi_{\pi\pi} \otimes \Phi_{\psi},
\end{eqnarray}
in which $\Phi_B$ and $\Phi_{\psi}$ are the $B$ meson  and charmonium DAs,
respectively. The two-pion DA $\Phi_{\pi\pi}$ collects the nonperturbative
dynamics in the $\pi\pi$ hadronization process. The hard kernel $H$, similar
to that in two-body decays, can be evaluated in perturbation theory.
The symbol $\otimes$ denotes the convolution in parton momenta
of all the perturbative and nonperturbative objects.

In this paper we will analyze the decays $B\rightarrow \psi (\pi\pi)_P$ with
%the meson $\psi$ encompassing the charmonia $J/\psi$ and $\psi(2S)$, and
the $P$-wave dipion system. We do not consider the corresponding decays of a
$B_s$ meson, in which the isovector resonant contributions are forbidden as
explained before. The decays $B_{(s)}\rightarrow J/\psi (\pi\pi)_S$ and
$B_{(s)}\rightarrow J/\psi (K\pi)_S$ as well as the $\psi(2S)$ counterparts, with
the $S$-wave $\pi\pi$ and $K\pi$ pairs, have been studied under the quasi-two-body
approximation in the PQCD approach \cite{prd91094024,epjc77199,prd97033006}.
The charmless $B$ meson decays into $P$-wave pion pairs in the longitudinal
polarization were investigated in Refs. \cite{plb76329,prd95056008,prd96036014}.
The three possible polarizations of the spin-1 $\psi$ meson
generate the longitudinal (0), parallel ($\parallel$), and perpendicular ($\perp$)
amplitudes, such that the two-pion DAs corresponding to both the longitudinal and
transverse polarizations are necessary nonperturbative inputs in our analysis.
We will include the two-pion $P$-wave DAs corresponding to the transverse polarization
into the PQCD formalism for the $B\rightarrow \psi (\pi\pi)_P$ decays.
It will be explained that the total momentum (angular momentum) of the pion pair
mimics the longitudinal (transverse) polarization of the $P$-wave dipion system.

The decomposition of the longitudinal two-pion DAs up to the twist-3 accuracy
has been presented in Ref. \cite{plb76329}, but that of the transverse DAs is not yet
available. Following the derivation in Refs. \cite{prd70054006,prd97034033},
the two-pion DAs can be parametrized in terms of the Gegenbauer polynomials
that depend on parton momentum fractions, and the Legendre polynomials that depend
on meson momentum fractions. Moreover, the two-pion DAs are normalized to the
time-like form factors, which contain both resonant and nonresonant
contributions to the dipion system. To be specific, we adopt the vector-dominance-model
parametrization for these form factors, which has been used to
fit the pion form factor measured via the $e^+e^-$ annihilation process
\cite{prd86032013}. Apart from the dominant $\rho(770)$ component, the two radial
excitations $\rho(1450)$ and $\rho(1700)$ as well as the $\rho$-$\omega$
interference effect were also taken into account.
Besides, the $B\rightarrow J/\psi \pi\pi$ modes are relevant to the determination
of the $CP$ violation phases in the $B$ system, which is, however, not the theme of
the present work. For recent progresses on this subject, refer to
\cite{plb719383,plb736186,prd91073007,plb713378,plb74238}.

The paper is organized as follows. In Sec.~\ref{sec:framework}
we define the involved kinematic variables and construct the two-pion DAs
for the longitudinal and transverse polarizations.
The numerical results are presented and discussed in Sec.~\ref{sec:results}.
The last section contains the conclusion. The factorization formulas for the
considered decay amplitudes are collected in the Appendix.

\section{ framework}\label{sec:framework}
\begin{figure}[!htbh]
\begin{center}
\vspace{1.5cm} \centerline{\epsfxsize=7 cm \epsffile{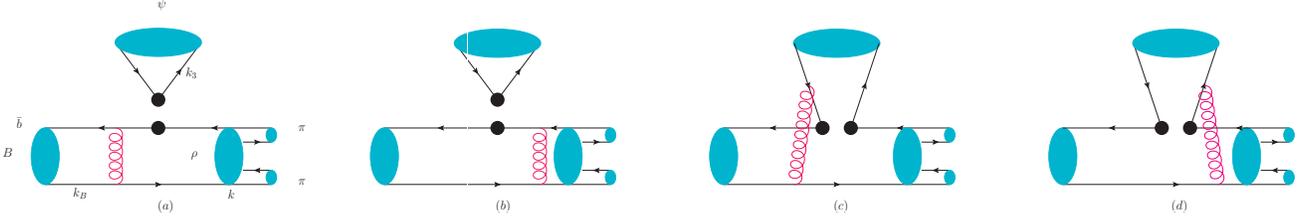}}
\caption{Leading-order Feynman diagrams for the
quasi-two-body decays  $B\rightarrow \psi \rho(\rightarrow \pi\pi)$, where
$\rho$ represents a $P$-wave $\pi\pi$ intermediate state, with
(a) and (b) the factorizable amplitudes, and (c) and (d) the
nonfactorizable amplitudes. }
 \label{fig:femy}
\end{center}
\end{figure}

We begin with the parametrization of the kinematic variables involved in
the decay $B (P_B)\rightarrow \psi (P_3) (\pi\pi)(P)$.
The momenta in the light-cone coordinates are chosen as
\begin{eqnarray}
 P_B&=&\frac{M}{\sqrt{2}}(1,1,\textbf{0}_{T}),
 \quad P_3=\frac{M}{\sqrt{2}}(r^2,1-\eta,\textbf{0}_{T}),
 \quad  P=\frac{M}{\sqrt{2}}(1-r^2,\eta,\textbf{0}_{T}),
\end{eqnarray}
in the $B$ meson rest frame,
with the mass ratio $r=m/M$, $m(M)$ being the charmonium ($B$ meson) mass, and
the variable $\eta=\omega^2/(M^2-m^2)$, $\omega^2 =P^2$ being the invariant mass squared
of the pion pair.  The momenta $p_1$ and $p_2$ of the two pions, obeying $p_1+p_2=P$,
are defined as
\begin{eqnarray}\label{eq:p1p2}
 p_1=(\zeta P^+, (1-\zeta)\eta P^+, \sqrt{\zeta(1-\zeta)}\omega,0),\quad
 p_2=((1-\zeta) P^+, \zeta\eta P^+, -\sqrt{\zeta(1-\zeta)}\omega,0),
\end{eqnarray}
with the pion momentum fraction $\zeta$.
We focus on the kinematic configuration, where $p_1$ and $p_2$
are almost collimated to each other with small amount of relative transverse momenta.
The valence quark momenta labeled by $k_B$, $k_3$, and $k$ in Fig. \ref{fig:femy} (a)
are parametrized as
\begin{eqnarray}
  k_B&=&(0,\frac{M}{\sqrt{2}}x_B,\textbf{k}_{BT}),\quad
  k_3=(\frac{M}{\sqrt{2}}r^2x_3,\frac{M}{\sqrt{2}}(1-\eta)x_3,\textbf{k}_{3T}),\quad
  k=(\frac{M}{\sqrt{2}}z(1-r^2),0,\textbf{k}_{T}),
\end{eqnarray}
in which $x_B$, $x_3$, $z$ denote the longitudinal momentum fractions,
and $k_{iT}$ represent the transverse momenta.

The hadronic matrix element for the $B$ meson is written as \cite{ppnp5185}
\begin{eqnarray}
\Phi_{B}(x,b)=\frac{i}{\sqrt{2N_c}}[(\rlap{/}{p_B}+M)\gamma_5\phi_{B}(x,b)],
\end{eqnarray}
with the impact parameter $b$ conjugate to the transverse momentum $\textbf{k}_{BT}$,
and the number of colors $N_c$. The $B$ meson DA $\phi_{B}(x,b)$ is
the same as in Refs. \cite{ppnp5185,prd65014007},
\begin{eqnarray}
\phi_{B}(x,b)=N x^2(1-x)^2\exp\left(-\frac{x^2M^2}{2\omega^2_b}-\frac{\omega^2_bb^2}{2}\right),
\end{eqnarray}
where the shape parameter $\omega_b=0.40\pm 0.04$ GeV has been fixed in
the study of the $B$ meson transition form factors \cite{plb5046,prd63054008},
and the coefficient $N$ is determined by the normalization $\int_0^1 dx \phi_{B}(x,b=0)=1$.

The hadronic matrix elements for the longitudinally and transversely polarized vector
charmonia are decomposed into
\begin{eqnarray}
\Phi_{\psi}^L&=&\frac{1}{\sqrt{2N_c}}[m \rlap{/}{\epsilon_{3L}}
\psi^L (x_3,b_3)+\rlap{/}{\epsilon_{3L}} \rlap{/}{p_3}\psi^t (x_3,b_3)],\nonumber\\
\Phi_{\psi}^T&=&\frac{1}{\sqrt{2N_c}}[m \rlap{/}{\epsilon_{3T}}
\psi^V (x_3,b_3)+\rlap{/}{\epsilon_{3T}} \rlap{/}{p_3}\psi^T (x_3,b_3)],
\end{eqnarray}
respectively, with the longitudinal and  transverse polarization vectors
\begin{eqnarray}
\epsilon_{3L}=\frac{1}{\sqrt{2(1-\eta)}r}(-r^2,1-\eta, \textbf{0}_{T}),
\quad \epsilon_{3T}=(0,0, \textbf{1}_{T}).
\end{eqnarray}
The explicit expressions of $\psi^i$ are referred to our previous
works \cite{prd90114030,epjc75293}.

The two-pion DAs can be related to the pion DAs through a perturbative
evaluation of the matrix elements \cite{prd70054006,prd97034033},
 \begin{eqnarray}\label{eq:matrix}
\langle \pi(p_1)\pi(p_2)|\bar{q}'(y^-)\Gamma q(0)|0\rangle,
\end{eqnarray}
as a timelike dipion production process, where $\Gamma$ denotes the
possible spin projectors $I$, $\gamma_5$,
$\gamma_{\mu}$, $\gamma_{\mu}\gamma_5$, $\sigma_{\mu\nu}$, and
$\sigma_{\mu\nu}\gamma_5$. The complete set of pion meson DAs up to
twist 3 is given by
\begin{eqnarray}
\Phi_{P_1}(p_1,x_1)&=&\frac{i}{\sqrt{2N_c}}\gamma_5[\rlap{/}{p_1}\phi_{P_1}^A(x_1)
+m_0\phi_{P_1}^P(x_1)+m_0(\frac{\rlap{/}{p_1}\rlap{/}{v}_B}{p_1\cdot v_B}-1)\phi_{P_1}^T(x_1)], \nonumber\\
\Phi_{P_2}(p_2,x_2)&=&\frac{i}{\sqrt{2N_c}}\gamma_5[\rlap{/}{p_2}\phi_{P_2}^A(x_2)
+m_0\phi_{P_2}^P(x_2)+m_0(\frac{\rlap{/}{p_2}\rlap{/}{v}_B}{p_2\cdot v_B}-1)\phi_{P_2}^T(x_2)],
\label{pion}
\end{eqnarray}
with the chiral scale $m_0$. The above decompositions, in which the $B$ meson
four-velocity $v_B=(1,0,0,0)$ is invariant under the frame rotation, hold for
the pion momenta $p_1$ and $p_2$ in arbitrary directions.
It is easy to see that the third structure in Eq.~(\ref{pion})
approaches to the conventional one in \cite{prd65014007},
\begin{eqnarray}
\frac{\rlap{/}{p_1}\rlap{/}{v}_B}{p_1\cdot v_B}\to \rlap{/}{n}_+ \rlap{/}{n}_-,
\end{eqnarray}
as $p_1$ is aligned with the plus direction $n_+ = (1,0,0_T)$, where
the dimensionless vector $n_-=  (0,1,0_T)$ is along the direction of the displacement
between the quarks $q$ and $q'$ in Eq.~(\ref{eq:matrix}).

The key to construct the transverse polarization vector $\epsilon_{T\mu}$
for the dipion system in terms of the kinematic variables in Eq.~(\ref{pion})
is to relate it to the orbital angular momentum
\begin{eqnarray}\label{trans}
\epsilon_{T\mu} \propto \epsilon_{\mu\nu\rho\sigma}
                            p_{1}^{\nu} p_{2}^{\rho} v_{B}^{\sigma},
\end{eqnarray}
with the Levi-Civita tensor $\epsilon_{\mu\nu\rho\sigma}$ under the
convention $\epsilon_{0123}=-1$.
The transverse polarization vector is then normalized into
\begin{eqnarray}\label{eq:jihua}
\epsilon_{T\mu} = \frac{\epsilon_{\mu\nu\rho\sigma}
                  p_1^{\nu} P^{\rho} n_{-}^{\sigma}}{\sqrt{\zeta(1-\zeta)}\omega P\cdot n_-}.
\end{eqnarray}
To arrive at the above expression, we have added $p_1^{\rho}$ to $p_2^{\rho}$
in Eq.~(\ref{trans}) to get the total momentum $P^\rho$ of the pion pair
without changing the result, and replaced $v_B$ by $n_-$,
because $P$ is dominated by the plus component.

Employing the pion DAs in Eq.~(\ref{pion}), adopting the definition
in Eq.~(\ref{eq:jihua}), and following the prescription in
\cite{prd70054006,prd97034033}, we obtain the nonlocal matrix elements in
Eq.~(\ref{eq:matrix}) for various spin projectors $\Gamma$ up to twist 3:
\begin{eqnarray}\label{eq:v}
\langle \pi \pi |\bar{q}'(y^-)\gamma_{\mu}q(0)|0\rangle&=&(2\zeta-1)P_{\mu}\int_0^1
dze^{izP\cdot y}\phi^0(z,\omega)\nonumber\\&&
-2\sqrt{\zeta(1-\zeta)}\omega \frac{\epsilon_{\mu\nu\rho\sigma}
\epsilon_{T}^{\nu}P^{\rho}n_{-}^{\sigma}}{P\cdot n_-}
\int_0^1 dze^{izP\cdot y}\phi^v(z,\omega),
\end{eqnarray}
\begin{eqnarray}\label{eq:s}
\langle \pi \pi |\bar{q}'(y^-)Iq(0)|0\rangle&=&\omega \int_0^1
dze^{izP\cdot y}\phi^s(z,\omega),
\end{eqnarray}
\begin{eqnarray}\label{eq:ta}
\langle \pi \pi |\bar{q}'(y^-)\sigma_{\mu\nu}\gamma_5q(0)|0\rangle
&=&-\sqrt{\zeta(1-\zeta)} \epsilon_{T\nu}P_{\mu} \int_0^1
dze^{izP\cdot y}\phi^T(z,\omega),
\end{eqnarray}
\begin{eqnarray}\label{eq:va}
\langle \pi \pi |\bar{q}'(y^-)\gamma_{\mu}\gamma_5q(0)|0\rangle
&=& i\sqrt{\zeta(1-\zeta)}\omega \epsilon_{T\mu} \int_0^1
dze^{izP\cdot y}\phi^a(z,\omega),
\end{eqnarray}
\begin{eqnarray}\label{eq:t}
\langle \pi \pi |\bar{q}'(y^-)\sigma_{\mu\nu}q(0)|0\rangle&=&-i
\frac{p_{1\mu}p_{2\nu}-p_{1\nu}p_{2\mu}}{\omega } \int_0^1
dze^{izP\cdot y}\phi^t(z,\omega),
\end{eqnarray}
\begin{eqnarray}\label{eq:p}
\langle \pi \pi |\bar{q}'(y^-)\gamma_5q(0)|0\rangle&=&0,
\end{eqnarray}
with the two-pion DAs $\phi^{0,T}$ and $\phi^{s,t,v,a}$ being of twist 2 and twist 3,
respectively.

Some detailed derivation of Eqs.~(\ref{eq:v})-(\ref{eq:p}) are outlined here.
For Eq. (\ref{eq:v}), we have applied the parametrizations for the
longitudinal and transverse components of $p_1-p_2$,
\begin{eqnarray}
(p_1-p_2)_{\mu}&\approx& (2\zeta-1)P_{\mu}, \nonumber\\
(p_1-p_2)^x &=&
%2\sqrt{\zeta(1-\zeta)}\omega=
-2\sqrt{\zeta(1-\zeta)}\omega\frac{\epsilon^{x\nu\rho\sigma}
                       \epsilon_{T\nu} P_{\rho} n_{-\sigma}}{P\cdot n_-},
\end{eqnarray}
where the $\zeta$-dependent factors will be absorbed into the corresponding two-pion DAs below.
The matrix element in Eq. (\ref{eq:ta}) for the choice $\mu,\nu=+,y$ is proportional to
\begin{eqnarray}
 \epsilon^{+y\rho\sigma} p_{1\rho} p_{2\sigma}=\epsilon^{\gamma y\rho\sigma} p_{1\rho} P_{\sigma}n_{-\gamma}
     =-\sqrt{\zeta(1-\zeta)}\omega P^+\epsilon_T^y ,
\end{eqnarray}
in which Eq.~(\ref{eq:jihua}) has been inserted.
It is pointed out that the structure $(p_{1\mu}p_{2\nu}-p_{1\nu}p_{2\mu})$ in Eq. (\ref{eq:t})
corresponds to $\rlap{/}{\epsilon}_L\rlap{/}{P}$ for the twist-3 DAs in
the longitudinally polarized pseudoscalar-vector meson pair \cite{prd70054006,prd97034033}.

We summarize the hadronic matrix elements $\Phi_{\pi\pi}^{L}$ ($\Phi_{\pi\pi}^{T}$)
for the pion pair associated with the longitudinal (transverse) polarization
from Eqs.~(\ref{eq:v})-(\ref{eq:t}) as
\begin{eqnarray}\label{eq:fuliye2}
\Phi_{\pi\pi}^{L}&=&\frac{1}{\sqrt{2N_c}}
[\rlap{/}{P}\phi^0(z,\zeta,\omega)+\omega \phi^s(z,\zeta,\omega)+
\frac{\rlap{/}{p}_1\rlap{/}{p}_2-\rlap{/}{p}_2\rlap{/}{p}_1}{\omega(2\zeta-1)}
\phi^t(z,\zeta,\omega)],\nonumber\\
\Phi_{\pi\pi}^{T}&=&\frac{1}{\sqrt{2N_c}}
[\gamma_5\rlap{/}{\epsilon}_T\rlap{/}{P} \phi^T(z,\zeta,\omega)
+\omega \gamma_5\rlap{/}{\epsilon}_T \phi^a(z,\zeta,\omega)
+i\omega\frac{\epsilon^{\mu\nu\rho\sigma}\gamma_{\mu}
\epsilon_{T\nu}P_{\rho}n_{-\sigma}}{P\cdot n_-} \phi^v(z,\zeta,\omega)],
\end{eqnarray}
where the projectors $\gamma_5\rlap{/}{\epsilon}_T\rlap{/}{P}$, $\gamma_5\rlap{/}{\epsilon}_T$,
and  $\epsilon^{\mu\nu\rho\sigma}\gamma_{\mu}
\epsilon_{T\nu}P_{\rho}n_{-\sigma}$ come from Eq.~(\ref{eq:ta}), Eq.~(\ref{eq:va}), and
the second line of Eq.~(\ref{eq:v}), respectively.
Our result for the longitudinal piece $\Phi_{\pi\pi}^{L}$ has the same form as in \cite{plb76329},
while the transverse one $\Phi_{\pi\pi}^{T}$ is new.
The two-pion DAs for various twists are expanded in
terms of the Gegenbauer polynomials, such as $C_2^{3/2}(1-2z)$:
 \begin{eqnarray}
\phi^0(z,\zeta,\omega)&=&\frac{3F^{\parallel}(\omega^2)}
{\sqrt{2N_c}}z(1-z)[1+a^0_2C_2^{3/2}(1-2z)](2\zeta-1),\nonumber\\
\phi^s(z,\zeta,\omega)&=&\frac{3F^{\perp}(\omega^2)}
{2\sqrt{2N_c}}(1-2z)[1+a_2^s(1-10z+10z^2)](2\zeta-1),\nonumber\\
\phi^t(z,\zeta,\omega)&=&\frac{3F^{\perp}(\omega^2)}
{2\sqrt{2N_c}}(1-2z)^2[1+a^t_2C_2^{3/2}(1-2z)](2\zeta-1),\nonumber\\
\phi^T(z,\zeta,\omega)&=&\frac{3F^{\perp}(\omega^2)}
{\sqrt{2N_c}}z(1-z)[1+a^{T}_2C_2^{3/2}(1-2z)]\sqrt{\zeta(1-\zeta)},\nonumber\\
\phi^a(z,\zeta,\omega)&=&\frac{3F^{\parallel}(\omega^2)}
{4\sqrt{2N_c}}(1-2z)[1+a_2^a(10z^2-10z+1)]\sqrt{\zeta(1-\zeta)},\nonumber\\
\phi^v(z,\zeta,\omega)&=&\frac{F^{\parallel}(\omega^2)}
{2\sqrt{2N_c}}\{\frac{3}{4}[1+(1-2z)^2]+a_2^v[3(2z-1)^2-1]\}\sqrt{\zeta(1-\zeta)},
\end{eqnarray}
in which we have introduced one Gegenbauer moment $a_2$ for each DA.
The decomposition of the above DAs is similar to that of the $\rho$ meson DAs, but
with the vector (tensor) decay constant $f_\rho$ ($f_\rho^T$) being replaced by the
timelike pion form factors $F^{\parallel}$ ($F^{\perp}$).

For the form factor $F^{\parallel}(\omega^2)$, we adopt the parametrization
in Ref.~\cite{prd86032013},
 \begin{eqnarray}\label{eq:fpi}
F^{\parallel}(\omega^2)=\left[BW^{GS}_{\rho}(\omega^2,m_{\rho},\Gamma_{\rho})
 \frac{1+c_{\omega}BW^{KS}_{\omega}(\omega^2,m_{\omega},\Gamma_{\omega})}{1+c_{\omega}}
 +\sum_{i}c_{i}BW^{GS}_{i}(\omega^2,m_{i},\Gamma_{i})\right]\left(1+\sum_{i}c_{i}\right)^{-1},
  \end{eqnarray}
with $i=\rho'(1450)$ and $\rho''(1700)$. The values of the masses $m_i$, the widths
$\Gamma_i$, the complex coefficients $c_i$, and the
BW functions of various resonances are referred to \cite{prd86032013}.
For the form factor $F^{\perp}(\omega^2)$, we employ the approximate relation
$F^{\perp}(\omega^2)/F^{\parallel}(\omega^2)\approx f_{\rho}^T/f_{\rho}$ for the
$\rho(770)$ resonance \cite{plb76329}.
Because the tensor decay constants $f^T$ for $\rho(1450)$ and $\rho(1700)$
are not known yet, we treat the corresponding modules $|c_i|$ in $F^\perp$ as free
parameters, but keep their phases the same as in \cite{prd86032013}.
The global fit to the existing data for the $B\rightarrow J/\psi \pi\pi$
branching ratios and polarization fractions \cite{prd90012003}
determines the central values of the dimensionless parameters
appearing in the two-pion DAs,
 \begin{eqnarray}\label{eq:para}
 a^0_2=0.2, \quad a_2^s=0.7, \quad a^t_2=-0.4,\quad  a^{T}_2=0.5,\quad
  a_2^a=0.4, \quad a_2^v=-0.5,\quad |c_{\rho'}|=0.316,\quad |c_{\rho''}|=0.272.
\end{eqnarray}

The differential branching fraction for the $B\rightarrow \psi \pi\pi$ decays into
$P$-wave pion pairs is expressed as
\begin{eqnarray}\label{eq:dfenzhibi}
\frac{d \mathcal{B}}{d \omega}=\frac{\tau \omega|\vec{p}_1||\vec{p}_3|}{32\pi^3M^3}
\sum_{i=0,\parallel,\perp}|\mathcal{A}_i|^2,
\end{eqnarray}
where the pion and charmonium three-momenta in the $\pi\pi$ center-of-mass frame are given by
\begin{eqnarray}
|\vec{p}_1|=\frac{\sqrt{\lambda(\omega^2,m_{\pi}^2,m_{\pi}^2)}}{2\omega}, \quad
|\vec{p}_3|=\frac{\sqrt{\lambda(M^2,m^2,\omega^2)}}{2\omega},
\end{eqnarray}
respectively, with the pion mass $m_{\pi}$ and  the K\"all\'en
function $\lambda (a,b,c)= a^2+b^2+c^2-2(ab+ac+bc)$.
The terms $\mathcal{A}_0$, $\mathcal{A}_{\parallel}$, and $\mathcal{A}_{\perp}$ represent
the longitudinal, parallel, and perpendicular polarization amplitudes in the transversity
basis, respectively.
The polarization fractions $f_{\lambda}$ with $\lambda=0$, $\parallel$,
and $\perp$ are then defined by
\begin{eqnarray}\label{pol}
f_{\lambda}=\frac{|\mathcal{A}_{\lambda}|^2}{|\mathcal{A}_0|^2
+|\mathcal{A}_{\parallel}|^2+|\mathcal{A}_{\perp}|^2}.
\end{eqnarray}

\section{Numerical results}\label{sec:results}

To proceed with the numerical analysis, we first
collect all the input quantities below.
The meson masses and the heavy quark masses take the central values
(in units of GeV) \cite{pdg2018}
\begin{eqnarray}
M&=&5.28, \quad m_b=4.8, \quad m_c=1.275, \quad m_{\rho}=0.775, \nonumber\\
m_{\pi^{\pm}}&=&0.140, \quad m_{\pi^{0}}=0.135, \quad m_{J/\psi}=3.097, \quad m_{\psi(2S)}=3.686.
\end{eqnarray}
The Cabibbo-Kobayashi-Maskawa (CKM) parameters in the Wolfenstein parametrization are set to
$\lambda = 0.22537$,  $A=0.814$,  $\bar{\rho}=0.117$, and $\bar{\eta}=0.355$ \cite{prd95056008}.
The decay constants (in units of GeV) and the $B$ meson lifetimes (in units of ps) are chosen
as \cite{prd95056008,prd90114030,epjc75293}
\begin{eqnarray}
f_B=0.19, \quad  f_{J/\psi}=0.405, \quad  f_{\psi(2S)}=0.296, \quad f_{\rho}=0.216,
 \quad f^T_{\rho}=0.184, \quad \tau_{B^0}=1.519,\quad \tau_{B^{\pm}}=1.638.
\end{eqnarray}
The resultant branching ratios $\mathcal{B}$ and the polarization fractions $f_{\lambda}$
together with the available experimental measurements from the LHCb Collaboration
for the $J/\psi$ involved modes are summarized in Table~\ref{tab:br}, and the
corresponding ones for $\psi(2S)$ are listed in Table \ref{tab:br1}.
Since the charged and neutral $B$ meson decays differ only in the lifetimes
and the isospin factor in our formalism, one can derive the branching ratios for
the $B^+$ meson by multiplying those for the $B^0$ meson by the ratio
$2\tau_{B^+}/\tau_{B^0}$.

The theoretical errors in Tables~\ref{tab:br} and \ref{tab:br1} are from some typical
sources, namely, the
two Gegenbauer moments in the twist-2 two-pion DAs, $a_2^0=0.2\pm 0.2$ and
$a_2^T=0.5\pm 0.5$, and the variation of the hard scales $t$ from $0.75t$ to $1.25t$,
which characterize the energy release in decay processes (see the factorization formulas
in the Appendix). It is worthwhile to mention that the  hard kernels are evaluated
only up to leading order plus the vertex corrections in this work,
so the theoretical accuracy still needs to be improved.
This is the case especially for $B$ meson decays into charmonia,
whose energy release may not be high enough for justifying the leading-order calculation.
It is then expected that the hadronic parameters extracted from the
data in the present framework should suffer larger theoretical uncertainty.
Therefore, we have considered a wide range for the
variation of the Gegenbauer moment $a_2^0 = 0.2\pm 0.2$, which covers
the central value $a_2^0=0.3$ extracted from the data for charmless $B$
meson decays in Ref. \cite{prd95056008}. Eventually, we will improve the accuracy of
our analysis and perform a global fit to all relevant data,
when determining the involved hadronic parameters.

One can see that the errors from the two Gegenbauer moments are comparable
and contribute to the major uncertainties as shown in
Tables \ref{tab:br} and \ref{tab:br1} , while the last one from the
hard scales is only of a few percents  due to the inclusion of the vertex
corrections. We have also examined the sensitivity of our results to the choice of other
Gegenbauer moments in the twist-3 two-pion DAs, $a_2^s$, $a_2^t$, $a_2^a$, and $a_2^v$,
in Eq.~(\ref{eq:para}). The first two give a comparable effect on the longitudinal
branching ratio as $a_2^0$ does. With the increase (decrease) of $a_2^s$ ($a_2^t$),
the total branching ratios and the longitudinal polarization factions become larger.
On the contrary, the last two have a little impact on the total branching ratios,
but can modify the relative importance of the parallel and perpendicular polarization
amplitudes. As we set $a_2^a=a_2^v=0$, the polarization fractions
$f_{\parallel}$ and $f_{\perp}$ are roughly equal. When $a_2^a$ and $a_2^v$ are
changed in the opposite direction, as indicated in Eq.~(\ref{eq:para}), the difference
between $f_{\parallel}$ and $f_{\perp}$ is enhanced and matches the data.
It can be understood from the  factorization formulas presented in the Appendix:
the contribution from $\phi^a$ to the parallel polarization amplitudes plays a role
similar to that from $\phi^v$ to the perpendicular polarization amplitudes, so
the inputs of $a_2^a$ and $a_2^v$
opposite in sign increase the difference between the two amplitudes.
It is also found that the coefficients $|c_i|$ in $F_{\perp}$ cause a significant
effect on the branching ratios for the $\rho(1450)$ and $\rho(1700)$ channels.
The variation of $|c_i|$ by $20\%$ results in the change of the branching ratios
by $40\%\sim50\%$. The uncertainties from other parameters in our formalism,
such as the decay constants and the CKM matrix elements, are not
discussed here. The polarization fractions are not sensitive to these parameters,
because they mainly yield an overall effect, which cancels in the ratios
defined by Eq.~(\ref{pol}).

\begin{table}
\caption{ PQCD results for the branching ratios and  the polarization fractions
of  the $P$-wave resonance channels in the  $B^0\rightarrow J/\psi \pi^+\pi^-$ decay.
The theoretical errors are attributed to the variation of the Gegenbauer
moments $a_2^0$ and $a_2^T$, and the hard scales $t$, respectively.
The data are taken from \cite{prd87052001,prd90012003,plb74238},
where the first uncertainty is statistical and the second is systematic.
The uncertainties from \cite{plb74238} are statistical only. }
\label{tab:br}
\begin{tabular}[t]{lcccc}
\hline\hline
$R$ & $\mathcal{B}(B^0\rightarrow J/\psi R (\rightarrow \pi^+\pi^-)) $ & $f_0(\%)$ & $f_{\parallel}(\%)$ &$f_{\perp}(\%)$\\
\hline
$\rho(770)$ & $(2.58^{+0.27+0.53+0.06}_{-0.25-0.38-0.04})\times 10^{-5}$ %{\color{red} ($2.71\times 10^{-5}$)}
& $57.9^{+4.0+10.1+0.6}_{-4.5-9.7-1.5}$ %{\color{red} ($59.9$)}
&$22.9^{+2.4+5.3+0.5}_{-2.2-6.0-0.4}$ %{\color{red} ($21.7$)}
& $19.2^{+2.1+4.4+1.0}_{-1.8-4.1-0.2}$ \\ %{\color{red} ($18.3$)}  \\
LHCb \cite{prd87052001} & $(2.49^{+0.20+0.16}_{-0.13-0.23})\times 10^{-5}$ &$\cdots$ &$\cdots$ & $\cdots$\\ %&{\color{blue}$63\pm4^{+6}_{-5} $}
LHCb \cite{prd90012003} & $(2.50\pm 0.10^{+0.18}_{-0.15})\times 10^{-5}$ & $57.4\pm0.2^{+1.3}_{-3.1}$
&$23.4\pm1.7^{+1.0}_{-1.3}$&$19.2\pm1.7^{+3.8}_{-1.2}$\\
LHCb \cite{plb74238}  & $(2.60\pm 0.10)\times 10^{-5}$ \footnotemark[1] & $56.7\pm1.8$ &$23.5\pm1.5$&$19.8\pm1.7$\\ \hline
$\rho(1450)$ & $(3.0^{+0.2+1.1+0.1}_{-0.1-0.6-0.0})\times 10^{-6}$
& $46^{+3+12+1}_{-1-11-4}$ &$29^{+1+9+2}_{-2-10-1}$& $25^{+1+3+1}_{-2-2-0}$\\
LHCb \cite{prd87052001} & $(2.1^{+1.0+2.2}_{-0.6-0.4})\times 10^{-6}$& $\cdots$& $\cdots$& $\cdots$\\ %& {\color{blue} $28^{+17+8}_{-13-12}$ }
LHCb \cite{prd90012003} & $(4.6\pm 1.1\pm 1.9)\times 10^{-6}$ & $58\pm10^{+14}_{-23}$
&$27\pm13^{+7}_{-11}$&$15\pm7^{+28}_{-10}$\\
LHCb \cite{plb74238}  & $(3.6\pm 0.7)\times 10^{-6}$ \footnotemark[1] & $47\pm11$ &$39\pm12$&$14\pm8$\\ \hline
$\rho(1700)$ & $(1.8^{+0.1+0.9+0.1}_{-0.0-0.5-0.0})\times 10^{-6}$
& $31^{+2+12+2}_{-0-9-0}$ &$38^{+0+9+0}_{-1-14-1}$& $31^{+1+2+1}_{-0-0-0}$\\
LHCb \cite{prd90012003} & $(2.0\pm 0.5\pm 1.2)\times 10^{-6}$ & $40\pm11^{+13}_{-23}$
&$24\pm14^{+7}_{-10}$&$36\pm14^{+28}_{-9}$\\
LHCb \cite{plb74238} & $(1.2\pm 0.3)\times 10^{-6}$ \footnotemark[1] & $29\pm12$ &$42\pm15$&$29\pm15$\\
\hline\hline
\end{tabular}
\footnotetext[1]{The fit fractions determined from the Dalitz plot analysis
have been converted into the branching fraction measurements.}
\end{table}

\begin{table}
\caption{ PQCD results for the branching ratios and  the polarization fractions of  the
$P$-wave resonance channels in the $B^0\rightarrow \psi(2S)  \pi^+\pi^-$ decay.
}
\label{tab:br1}
\begin{tabular}[t]{lcccc}
\hline\hline
$R$ & $\mathcal{B}(B^0\rightarrow \psi(2S) R(\rightarrow \pi^+\pi^-)) $ & $f_0(\%)$ & $f_{\parallel}(\%)$ &$f_{\perp}(\%)$\\
\hline
$\rho(770)$ & $(1.0^{+0.1+0.2+0.0}_{-0.1-0.2-0.0})\times 10^{-5}$
& $50^{+3+9+1}_{-2-8-0}$ &$26^{+1+5+0}_{-2-7-1}$& $24^{+1+3+0}_{-1-3-1}$ \\
$\rho(1450)$ & $(8.2^{+0.1+2.3+0.4}_{-0.0-1.5-0.2})\times 10^{-7}$
 & $46^{+1+11+3}_{-0-10-3}$ &$28^{+0+9+2}_{-1-10-2}$& $26^{+0+1+0}_{-0-1-1}$\\
\hline\hline
\end{tabular}
\end{table}

It is obvious that
both our branching ratio and three polarization fractions
for the $\rho(770)$ channel agree well with the high-precision LHCb
data \cite{prd87052001,prd90012003,plb74238} in Table \ref{tab:br}.
Although the central values of the measured branching ratios for the
$\rho(1450)$ resonance vary in a wide range $(2.1\sim 4.6)\times 10^{-6}$,
their PDG weighted average leads to $2.9^{+1.6}_{-0.7}\times 10^{-6}$ \cite{pdg2018},
in good consistency with our prediction. For the $\rho(1700)$ channel, the
LHCb Collaboration got $\mathcal{B}(B^0\rightarrow J/\psi \rho''
(\rightarrow \pi^+\pi^-))=(2.0\pm 0.5\pm 1.2)\times 10^{-6}$ \cite{prd90012003},
while the subsequent measurement gave $(1.2\pm 0.3)\times 10^{-6}$ \cite{plb74238}
with the statistical uncertainty only. Our prediction $(1.8^{+0.9}_{-0.5})\times 10^{-6}$
is in between, and matches both data within errors.
For the $\psi(2S)$ involved modes, although the LHCb Collaboration \cite{npb871403}
also observed a dominant contribution to the $B^0 \rightarrow \psi(2S) \pi^+\pi^-$
decay from  the $\rho(770)$ resonance, the detailed
partial wave analysis for determining its fraction is still missing
due to a limited number of events.

Summing over all the contributing $P$-wave resonances in the $\pi\pi$ invariant mass
spectra $[2m_{\pi}, M-m]$, we have the total branching ratios
\begin{eqnarray}\label{eq:p-wave}
\mathcal{B}(B^0\rightarrow J/\psi (\pi^+\pi^-)_P)&=&(3.1^{+0.4+0.8+0.2}_{-0.2-0.5-0.0})\times 10^{-5},\nonumber\\
\mathcal{B}(B^0\rightarrow \psi(2S) (\pi^+\pi^-)_P)&=&(1.2^{+0.1+0.3+0.0}_{-0.1-0.2-0.0})\times 10^{-5},
\end{eqnarray}
where the sources of the errors have been interpreted before.
The former amounts up to $78\%$ of the total three-body
branching ratio $\mathcal{B}(B^0\rightarrow J/\psi \pi^+\pi^-)=(3.96\pm0.17)\times 10^{-5}$ \cite{pdg2018}.
As noticed in \cite{prd87052001}, the $S$-wave $f_0(500)$ and $D$-wave $f_2(1270)$
resonances, besides the $P$-wave ones, were also produced significantly in the
 $J/\psi \pi^+\pi^-$ final states.
The best fit model in \cite{prd87052001} implies that  %$\pm 149$ MeV
one full $\rho(770)$ meson width contains $11.9\%$ $S$-wave component and
$0.72\%$ $D$-wave component. Therefore, it is reasonable to leave the
remaining $22\%$ to the $S$-wave and $D$-wave contributions, as well as the nonresonant
one and their interference in the entire invariant mass range.
We estimate from Eq. (\ref{eq:p-wave}) the ratio of the branching fractions,
\begin{eqnarray}
\frac{\mathcal{B}(B^0\rightarrow \psi(2S) (\pi^+\pi^-)_P)}
{\mathcal{B}(B^0\rightarrow J/\psi (\pi^+\pi^-)_P)}=0.39^{+0.01}_{-0.03},
\end{eqnarray}
in which all the uncertainties have been added in quadrature.
The value is slightly lower than the LHCb  measurement \cite{npb871403}
\begin{eqnarray}
\frac{\mathcal{B}(B^0\rightarrow \psi(2S) \pi^+\pi^-)}
{\mathcal{B}(B^0\rightarrow J/\psi \pi^+\pi^-)}=0.56\pm
0.07(\text{stat})\pm0.05 (\text{syst})\pm0.01(\mathcal{B}),
\end{eqnarray}
where the third uncertainty corresponds to the one from
the dilepton branching fractions of the $J/\psi$ and $\psi(2S)$ charmonium decays.
The minor discrepancy may be resoled by including other partial wave contributions.

The resonant decay rate obeys a simple factorization relation under the
narrow width approximation,
\begin{eqnarray}
\mathcal{B}(B^0\rightarrow \psi R(\rightarrow \pi^+\pi^-))=
\mathcal{B}(B^0\rightarrow \psi R)\mathcal{B}(R \rightarrow \pi^+\pi^-),
\end{eqnarray}
from which we extract the two-body $B\rightarrow \psi R$ branching ratios,
given the input of $\mathcal{B}(R \rightarrow \pi^+\pi^-)$.
Combining the experimental fact  $\mathcal{B}(\rho\rightarrow \pi\pi) \sim 100\%$ \cite{pdg2018}
and the estimates of $\mathcal{B}(\rho'\rightarrow \pi\pi)=10.04^{+5.23}_{-2.61}\%$
and $\mathcal{B}(\rho''\rightarrow \pi\pi)=8.11^{+2.22}_{-1.47}\%$ in Ref.~\cite{prd96036014},
we obtain the central values
\begin{eqnarray}
\mathcal{B}(B^0\rightarrow J/\psi \rho)&=&2.58\times 10^{-5},\nonumber\\
\mathcal{B}(B^0\rightarrow J/\psi \rho')&=&3.0\times 10^{-5},\nonumber\\
\mathcal{B}(B^0\rightarrow J/\psi \rho'')&=&2.2\times 10^{-5},\nonumber\\
\mathcal{B}(B^0\rightarrow \psi(2S) \rho)&=&1.0\times 10^{-5},\nonumber\\
\mathcal{B}(B^0\rightarrow \psi(2S) \rho')&=&8.2\times 10^{-6}.
\end{eqnarray}
It is seen that both $\mathcal{B}(B^0\rightarrow J/\psi \rho)$
and $\mathcal{B}(B^0\rightarrow \psi(2S) \rho)$ are consistent with those
derived in the PQCD framework for two-body decays \cite{epjc77610}.

\begin{figure}[tbp]
\begin{center}
\setlength{\abovecaptionskip}{0pt}
\centerline{
\hspace{4cm}\subfigure{\epsfxsize=13 cm \epsffile{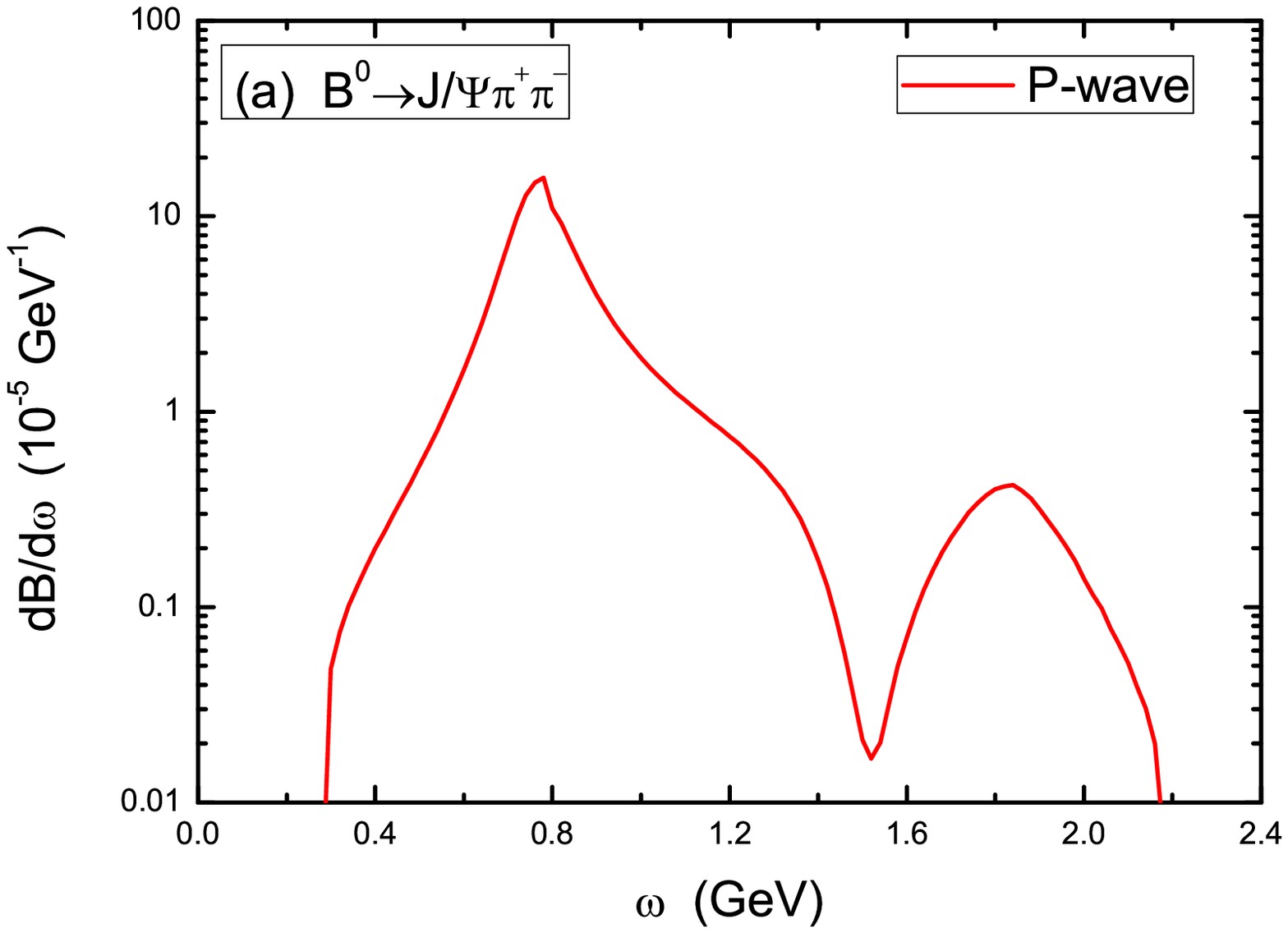} }
\hspace{-6cm}\subfigure{ \epsfxsize=13 cm \epsffile{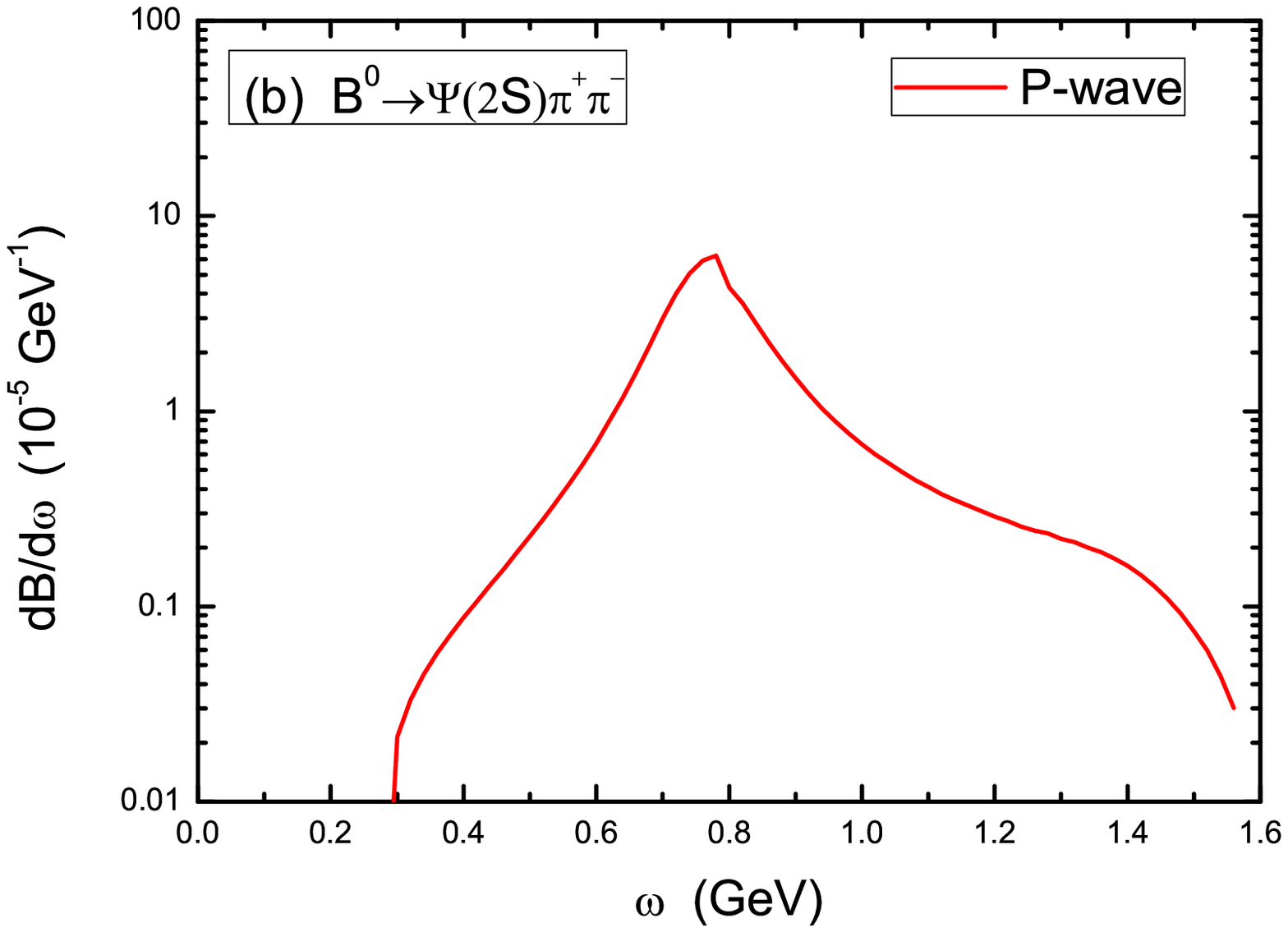}}}
\vspace{-3cm}\caption{$P$-wave contribution to the differential branching fractions of
the modes (a) $B^0\rightarrow J/\psi \pi^+\pi^-$ and (b) $B^0\rightarrow \psi(2S) \pi^+\pi^-$.}
 \label{fig:pwave}
\end{center}
\end{figure}
\begin{figure}[tbp]
\begin{center}
\setlength{\abovecaptionskip}{0pt}
\centerline{
\hspace{4cm}\subfigure{\epsfxsize=13 cm \epsffile{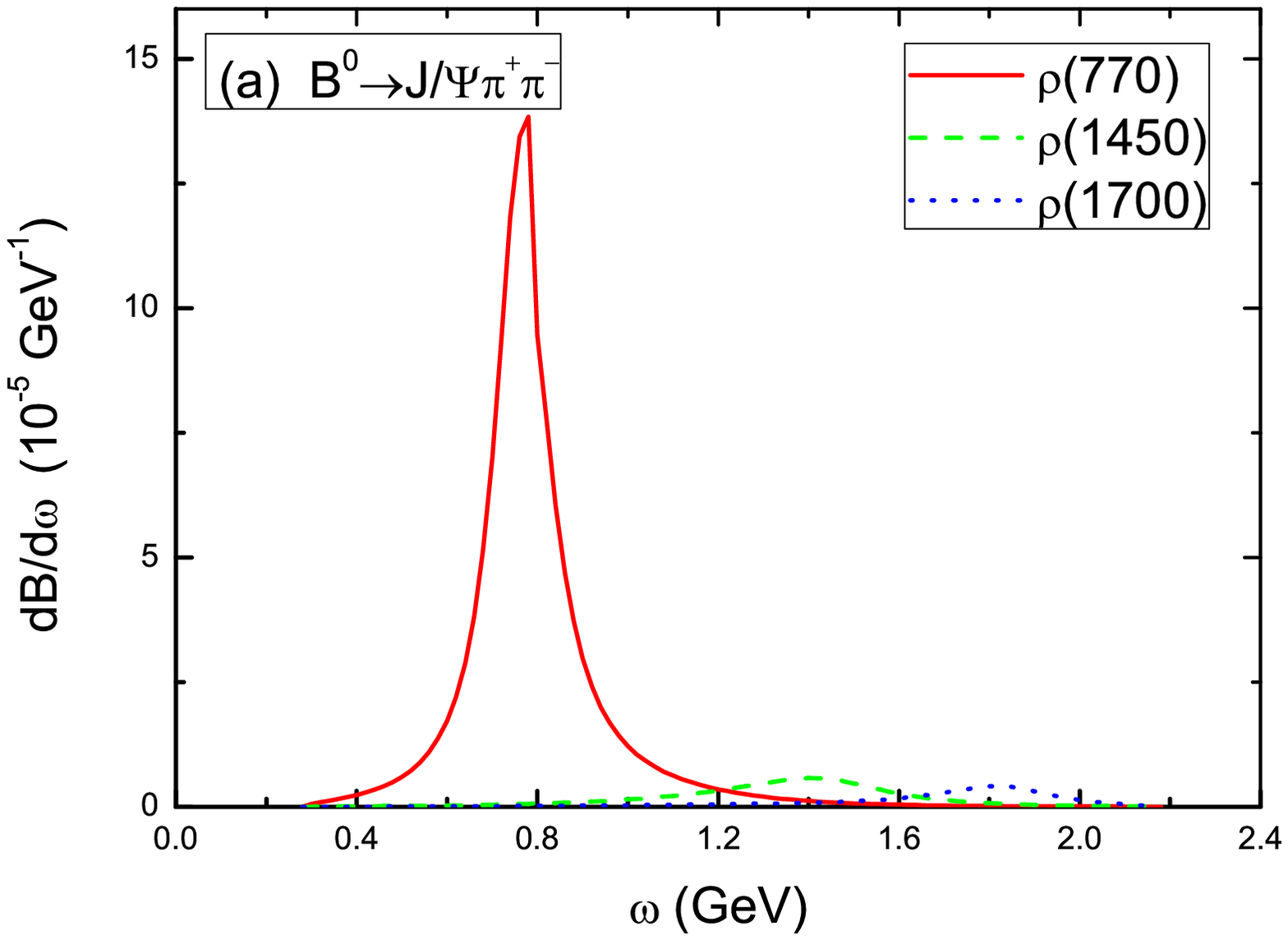} }
\hspace{-6cm}\subfigure{ \epsfxsize=13 cm \epsffile{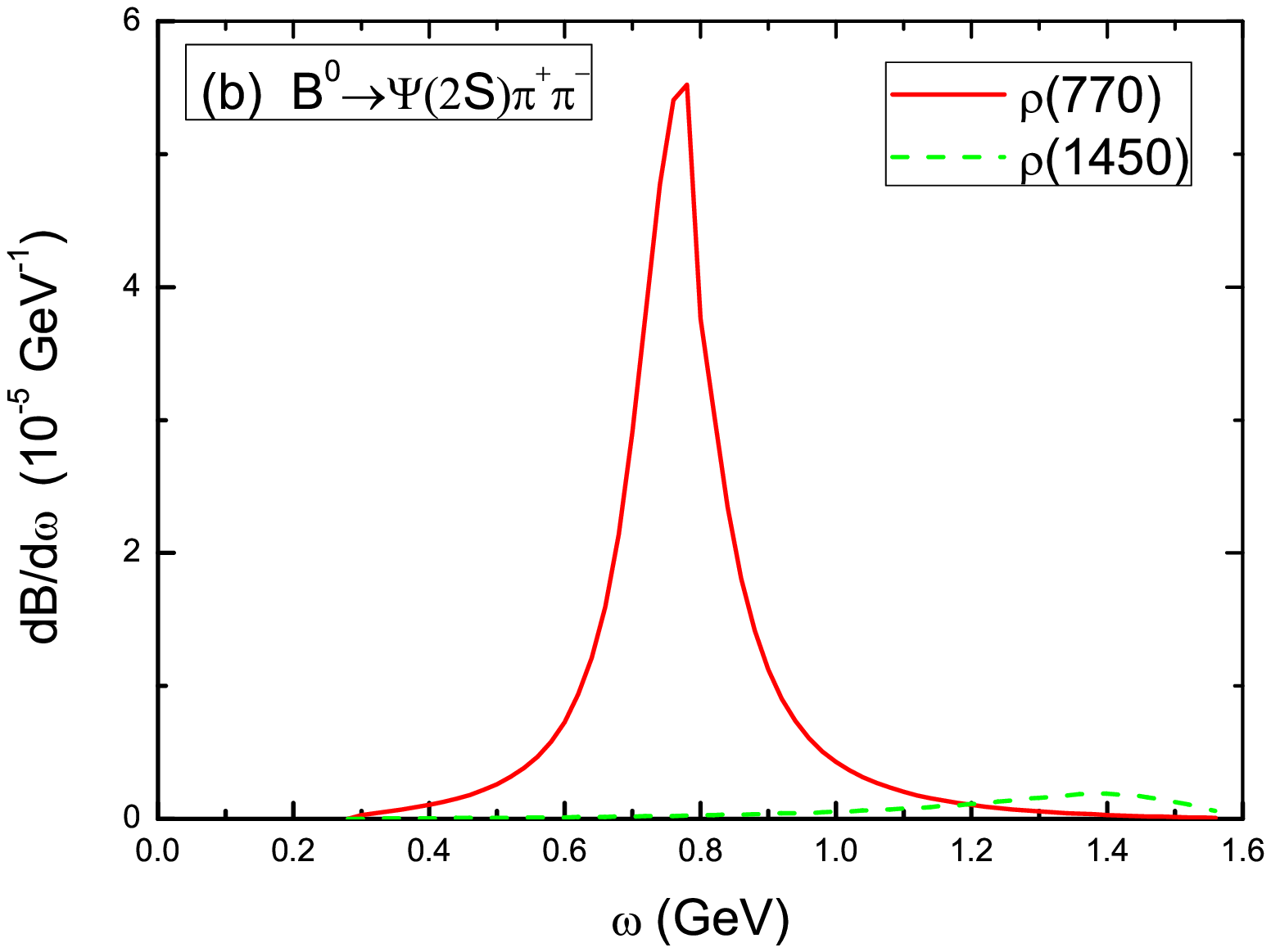}}}
\vspace{-3cm}\caption{$\rho(770)$, $\rho(1450)$, and $\rho(1700)$
resonance contributions to the differential branching fractions of
(a) $B^0\rightarrow J/\psi \pi^+\pi^-$ and (b) $B^0\rightarrow \psi(2S) \pi^+\pi^-$,
which are displayed by the solid red, dashed green, and dotted blue curves, respectively. }
 \label{fig:reso}
\end{center}
\end{figure}

We plot in Fig.~\ref{fig:pwave} the total differential branching fractions
in the $P$-wave $\pi^+\pi^-$ invariant mass for  the considered decays. The curve for
the $B^0\rightarrow J/\psi \pi^+\pi^-$ mode is similar to those for the
charmless $B\rightarrow P \pi\pi$ decays \cite{prd96036014},
since the same time-like form factors for the two-pion DAs, fitted by the
$BABAR$ Collaboration via the $e^+e^-$ annihilation
process \cite{prd86032013}, have been adopted.
One finds a dip appearing at the invariant mass
around $1.5 \sim  1.6$ GeV in Fig. \ref{fig:pwave}(a), that is usually interpreted
as the destructive interference between the $\rho(1450)$ and $\rho(1700)$
channels \cite{prd86032013,prd78072006}. In fact, the best fit model also shows
that the destructive interference between $\rho(1450)$ and $\rho(1700)$ is
comparable with their individual fit fractions (see Tables VII and IX in Ref.~\cite{prd90012003}).
However, the dip is not observed in Fig. \ref{fig:pwave}(b), because the $\rho(1700)$ state is
beyond the dipion invariant mass spectra for the $B^0\rightarrow \psi(2S) \pi^+\pi^-$ mode.
Both cases exhibit a clear $\rho$-$\omega$ interference pattern in the $\rho$ peak region.
The individual resonance contributions are displayed in Fig.~\ref{fig:reso},
where the red solid, green dashed, and blue dotted curves represent
those from $\rho(770)$, $\rho(1450)$, and $\rho(1700)$, respectively.
The different shapes among these individual channels are mainly governed
by the corresponding BW functions and parameters $c_i$ in Eq.~(\ref{eq:fpi}).
As expected, the $\rho(770)$ production is apparently dominant.
Comparing Tables \ref{tab:br} and \ref{tab:br1} with Eq.~(\ref{eq:p-wave}),
the $\rho(770)$ resonance accounts for $83\%$ of the total $P$-wave branching fractions
in both the $B^0\rightarrow J/\psi \pi^+\pi^-$ and $\psi(2S) \pi^+\pi^-$
decays, while the higher $\rho(1450)$ and $\rho(1700)$ resonances contribute
less than $10\%$. The obtained distributions in the $P$-wave $\pi\pi$ mass as well as
the individual resonance contributions agree fairly well with the LHCb data
shown in Fig. 13 of Ref.~\cite{prd90012003} and in Fig. 4 of Ref.~\cite{npb871403}.

\section{ conclusion}\label{sec:sum}
In this paper we have performed the analysis of the $B\rightarrow \psi \pi\pi$ decays
under the quasi-two-body approximation in the PQCD framework by introducing the
two-pion DAs. Since both the charmonium and the $P$-wave pion pair
in the final state carry the spin degrees of freedom, the two-pion DAs corresponding to
both the longitudinal and transverse polarizations are the necessary nonperturbative inputs,
and were constructed through a perturbative evaluation of the associated hadronic matrix
elements as a timelike process. It was observed that the total
momentum and the orbital angular momentum of the $P$-wave dipion system mimics its longitudinal
and transverse polarizations, respectively. The two-pion DAs for various
spin projectors were then decomposed in terms of the Gegenbauer polynomials
that depend on parton momentum fractions, and the Legendre polynomials that depend
on meson momentum fractions up to twist 3. The timelike form factors, normalizing
the two-pion DAs, were parametrized to consist of a
linear combination of the $\rho$,  $\rho'$, and $\rho''$ resonant contributions
together with the $\rho$-$\omega$ interference.

We have determined the hadronic parameters involved in the two-pion DAs from a
global fit to the data of the $B^0\rightarrow J/\psi \rho(\rightarrow \pi^+\pi^-)$
branching ratios and polarization fractions with good consistency. In particular,
the resultant differential branching fractions in the $P$-wave dipion invariant mass
and individual resonance contributions match the LHCb data.
We have also predicted the branching ratios and the
polarization fractions of the $B^0\rightarrow \psi(2S)\rho(\rightarrow  \pi^+\pi^-)$
decays, which can be confronted with future measurements.
As a by-product, we extracted the two-body $B^0\rightarrow \psi \rho$
branching ratios from the results for the corresponding quasi-two-body modes
by employing the narrow width approximation. The predictions for the $\rho(770)$
channels are in accordance with our previous PQCD calculations performed for
two-body decays. The consistency between the
three-body and two-body analyses supports the PQCD approach to exclusive
charmonium $B$ meson decays. The predictions for the higher excited intermediate states
still need to be tested at the ongoing and forthcoming experiments.

\begin{acknowledgments}
We acknowledge Wen-Fei Wang and Chao Wang for helpful discussions. This work was
supported in part by the National Natural Science Foundation of China
under Grants No.11605060 and No.11547020, by the Program for the Top Young
Innovative Talents of Higher Learning Institutions of Hebei Educational
Committee under Grant No. BJ2016041, and the Ministry of Science and Technology
of R.O.C. under Grant No. MOST-107-2119-M-001-035-MY3.
\end{acknowledgments}

\begin{appendix}
\section{decay amplitudes}\label{sec:app}

Before presenting the explicit factorization formula for each
$B^0\rightarrow \psi \pi\pi$ decay amplitude in this appendix,
we make a remark on the factorization theorem for
hadronic $B$ meson decays into charmonia. It has been
argued  \cite{prd63074011} that the QCD factorization (QCDF) approach is
applicable to exclusive $B$ meson decays into $J/\psi$, since the
transverse size of $J/\psi$ becomes small in the heavy quark limit.
On the other hand, the $k_T$ factorization theorem also holds for $B$
meson decays containing charmonia in the heavy quark limit under the power
counting $m_c/m_b$, $\Lambda_{\rm QCD}/m_c \ll 1$, with the QCD scale
$\Lambda_{\rm QCD}$, as elaborated in \cite{prd67054028}. Because we
focus on the resonant region of the dipion system, what we
studied here are basically quasi-two-body decays, and the
reasoning in \cite{prd67054028} for their factorization still applies.
That is, the PQCD approach is expected to be suitable for describing
the $B^0\rightarrow \psi \pi\pi$ decays.

The contributions from
the longitudinal polarization, the normal polarization, and the transverse
polarization are labelled by the subscripts $L$, $N$ and $T$, respectively.
The contributions from the $(V-A)\otimes(V-A)$,
$(V-A)\otimes(V+A)$, and $(S-P)\otimes(S+P)$ operators are labelled by
the superscripts $LL$, $LR$, and $SP$, respectively.
The total decay amplitude is decomposed into
\begin{eqnarray}\label{eq:alnt}
\mathcal{A}=\mathcal{A}_L+\mathcal{A}_N \epsilon_{T}\cdot \epsilon_{3T}
+i \mathcal{A}_T \epsilon_{\alpha\beta\rho\sigma} n_+^{\alpha} n_-^{\beta} \epsilon_{T}^{\rho} \epsilon_{3T}^{\sigma},
\end{eqnarray}
where the three individual polarization amplitudes are written as
\begin{eqnarray}
\mathcal{A}_{L,N,T}&=&\frac{G_F}{\sqrt{2}}\Big\{V^*_{cb}V_{cs}
\Big [(C_1+\frac{1}{3}C_2)\mathcal{F}_{L,N,T}^{LL}+C_2\mathcal{M}_{L,N,T}^{LL} \Big]
-V^*_{tb}V_{ts}\Big [(C_3+\frac{1}{3}C_4+C_9+\frac{1}{3}C_{10})\mathcal{F}_{L,N,T}^{LL}+\nonumber\\
&&(C_5+\frac{1}{3}C_6+C_7+\frac{1}{3}C_{8})\mathcal{F}_{L,N,T}^{LR}
+(C_4+C_{10})\mathcal{M}_{L,N,T}^{LL}+(C_6+C_8)\mathcal{M}_{L,N,T}^{SP}\Big ]\Big\},
\end{eqnarray}
with the CKM matrix elements $V_{ij}$ and the Fermi coupling constant $G_F$.
The Wilson coefficients $C_i$ encode the hard dynamics of weak decays.
The above amplitudes are related to those in Eq.~(\ref{eq:dfenzhibi}) via
\begin{eqnarray}
\mathcal{A}_0=\mathcal{A}_L, \quad \mathcal{A}_{\parallel}=\sqrt{2}\mathcal{A}_{N},
\quad \mathcal{A}_{\perp}=\sqrt{2}\mathcal{A}_{T}.
\end{eqnarray}

The explicit amplitudes $\mathcal{F(M)}$ from the factorizable (nonfactorizable)
diagrams in Fig. \ref{fig:femy} read as
\begin{eqnarray}
\mathcal{F}^{LL}_L&=&\frac{8\pi C_F f_{\psi} M^4}{\sqrt{1-\eta}} \int_0^1dx_B dz \int_0^{\infty} b_B db_Bbdb\phi_B(x_B,b_B)\nonumber\\&&
\{[-\phi^0(r^2 (-2 \eta  z+2 z+1)+(\eta -1) (z+1))- \sqrt{\eta(1-r^2)}(\phi^s(\eta +r^2 (2 (\eta -1) z+1)-2 \eta  z+2 z-1)\nonumber\\&&
+\phi^t(\eta +r^2 (2 (\eta -1) z-1)-2 \eta  z+2 z-1))]E_e(t_a)h_a(x_B,z,b_B,b)\nonumber\\&&+[2\phi^s(\sqrt{\eta(1-r^2)}(-\eta +r^2 x_B-r^2+1))
-\phi^0(-\eta ^2+\eta +\eta ^2 r^2-2 \eta  r^2+r^2 x_B)]E_e(t_b)h_b(x_B,z,b_B,b)\},\nonumber\\
& &
\end{eqnarray}
\begin{eqnarray}
\mathcal{M}^{LL}_L&=&-\frac{32\pi C_F  M^4}{\sqrt{6(1-\eta)\eta(1-r^2) }} \int_0^1dx_B dz dx_3\int_0^{\infty} b_B db_Bb_3db_3\phi_B(x_B,b_B)\nonumber\\&&[\phi^0(\eta+r^2-1)\sqrt{\eta(1-r^2)}
-2\eta(r^2-1)\phi^t]\nonumber\\&&[r^2\psi^L(2(\eta-1)x_3+x_B-\eta z+z)-2(\eta-1)rr_c\psi^t+(\eta-1)z\psi^L]E_n(t_d)h_d(x_B,z,x_3,b_B,b_3),
\end{eqnarray}
\begin{eqnarray}
\mathcal{F}^{LL}_N&=&8\pi C_F f_{\psi} M^4r \int_0^1dx_B dz \int_0^{\infty} b_B db_Bbdb\phi_B(x_B,b_B)\nonumber\\&&
\{[\sqrt{\eta(1-r^2)}(\phi^a(r^2z-z-2)-(r^2-1)z\phi^v)
+\phi^T(r^2-1+\eta(-2r^2z+2z-1))]E_e(t_a)h_a(x_B,z,b_B,b)
\nonumber\\&&+\sqrt{\eta(1-r^2)}[\phi^a(-\eta+r^2+x_B-1)+
\phi^v(\eta+r^2-x_B-1)]E_e(t_b)h_b(x_B,z,b_B,b)\},
\end{eqnarray}
\begin{eqnarray}
\mathcal{F}^{LL}_T&=&8\pi C_F f_{\psi} M^4r \int_0^1dx_B dz \int_0^{\infty} b_B db_Bbdb\phi_B(x_B,b_B)\nonumber\\&&
\{[\sqrt{\eta(1-r^2)}(\phi^v(r^2z-z-2)-(r^2-1)z\phi^a)
+\phi^T(r^2-1-\eta(-2r^2z+2z-1))]E_e(t_a)h_a(x_B,z,b_B,b)
\nonumber\\&&+\sqrt{\eta(1-r^2)}[\phi^v(-\eta+r^2+x_B-1)+
\phi^a(\eta+r^2-x_B-1)]E_e(t_b)h_b(x_B,z,b_B,b)\},
\end{eqnarray}
\begin{eqnarray}
\mathcal{M}^{LL}_N&=&-\frac{64\pi C_F  M^4}{\sqrt{6}}
\int_0^1dx_B dz dx_3\int_0^{\infty} b_B db_Bb_3db_3\phi_B(x_B,b_B)\nonumber\\
&&\{\phi^T[r\psi^V(-\eta x_3+x_3-x_B+\eta z)+(\eta-1)r_c\psi^T]-\nonumber\\
&&\sqrt{\eta(1-r^2)}\phi^a[r\psi^V(-\eta x_3+x_3-x_B+z)+(\eta-1)r_c\psi^T]\}E_n(t_d)h_d(x_B,z,x_3,b_B,b_3),
\end{eqnarray}
\begin{eqnarray}
\mathcal{M}^{LL}_T&=&-\frac{64\pi C_F  M^4}{\sqrt{6}}
\int_0^1dx_B dz dx_3\int_0^{\infty} b_B db_Bb_3db_3\phi_B(x_B,b_B)\nonumber\\
&&\{\phi^T[r\psi^V(-\eta x_3+x_3-x_B-\eta z)+(\eta-1)r_c\psi^T]-\nonumber\\
&&\sqrt{\eta(1-r^2)}\phi^v[r\psi^V(-\eta x_3+x_3-x_B+z)+(\eta-1)r_c\psi^T]\}E_n(t_d)h_d(x_B,z,x_3,b_B,b_3),
\end{eqnarray}
\begin{eqnarray}
\mathcal{F}^{LR}_{L,N,T}&=&\mathcal{F}^{LL}_{L,N,T},
\end{eqnarray}
\begin{eqnarray}
\mathcal{M}^{SP}_{L,N,T}&=&-\mathcal{M}^{LL}_{L,N,T},
\end{eqnarray}
with $r_c=m_c/M$, $m_c$ being the charm quark mass, the color factor $C_F=4/3$,
and the decay constant $f_{\psi}$ of the charmonium.
The expressions for the evolution functions $E$, the
hard kernels $h$, and the hard scales $t_{a,b,c,d}$ can be found in the Appendix of
Ref.~\cite{prd91094024}. We point out that the amplitudes $\mathcal{F}$
correspond to the $B\to\pi\pi$ transition form factors, which have been computed
in QCD light-cone sum rules \cite{plb730336,Hambrock:2015aor}.

In addition, the vertex corrections to the factorizable
diagrams in Fig. \ref{fig:femy} are included through the modification
of the Wilson coefficients as done in the QCDF approach \cite{bbns1,bbns2,bbns3},
according to the argument in \cite{Li:2005kt}. Note that the first step
of the factorization of these diagrams is the same in the QCDF and PQCD
approaches, at which the Wilson coefficients are factorized out of the
exclusive $B$ meson decays. The difference of the two approaches stems
from whether the remaining hadronic matrix elements of effective
operators, namely, the soft form factors, are factorizable. Due to
the different power counting on parton transverse momenta, these soft
form factors are not factorizable in QCDF, but are in PQCD. Once the
factorization is established, one can calculate radiative corrections
to each involved piece separately. Since the Wilson
coefficients are the same in the two approaches, the vertex corrections
to this piece obtained in QCDF can be applied to PQCD.
Moreover, the infrared divergences in the vertex corrections cancel, when
they are summed over, as stated in Ref. \cite{prd63074011}.
Therefore, it is not necessary to introduce parton transverse momenta
into the evaluation of these corrections \cite{Li:2005kt}, and the
QCDF results can be adopted directly and consistently.

\end{appendix}

\end{document}